# Role of grain boundary and dislocation loop in H blistering in W: A Density functional theory assessment

W. Xiao and W. T. Geng[a]

*School of Materials Science & Engineering, University of Science & Technology Beijing, Beijing 100083, China,*

January 2, 2011

We report a first-principles density functional theory study on the role of grain boundary and dislocation loop in H blistering in W. At low temperature, the Σ3(111) tilt grain boundary, when combined with a vacancy of vanishing formation energy, can trap up to nine H atoms per (1×1) unit in (111) plane. This amount of H weakens the cohesion across the boundary to an extent that a cleavage along the GB is already exothermic. At high temperature, this effect can be still significant. For an infinitely large dislocation loop in (100) plane, four H can be trapped per (1×1) unit even above room temperature, incurring a decohesion strong enough to break the crystal. Our numerical results demonstrate unambiguously the grain boundaries and dislocation loops can serve as precursors of H blistering. In addition, no $H_2$ molecules can be formed in either environment before fracture of W bonds starts, well explaining the H blistering in the absence of voids during non-damaging irradiation.

---

[a] To whom correspondence should be addressed. E-mail: geng@ustb.edu.cn

Hydrogen blistering in tungsten, a promising candidate to serve as the plasma-facing material in fusion reactors, has been under intensive study recently.[1,2,3] A great amount of irradiation experiments have been carried out to shed light on the relationships of retention of H isotopes and the conditions of irradiation (energy, temperature, flux, and fluence) and material preparation (alloying additions, processing, and surface treatment). The blistering in single-crystalline W under non-damaging irradiation is generally believed to occur in association with a very high mobile H concentration as low temperature and long irradiation time are needed, but the precise picture of the spontaneous emission of vacancies in the very initial stage remains elusive.[4,5,6,7] By comparison, H blistering in polycrystalline W under high-energy flux is more easily to observe because (i) the pre-existed defects like grain boundaries and dislocation loops can help to build up a high local H concentration by trapping H atoms from the surrounding bulk environment and (ii) the micro-voids formed upon damaging irradiation can even accommodate a small amount $H_2$ molecules which can serve as immediate vanguards in gas driven fracture of the crystal lattice.[8,9,10]

The mechanism of H trapping at grain boundaries and dislocations is very similar to what happens at a vacancy, where low-electron-density provides a more comfortable environment for H than interstitial site in perfect lattice.[11] All these traps may act as seeds for bubble growth. Using density functional theory calculations, Liu *et al.*[12] have demonstrated that at zero temperature a vacancy can trap up to ten H atoms until a $H_2$ molecule is formed inside the vacancy. Using the same code, Ohsawa *et al.*[13] have shown that a vacancy can trap up to 12 H atoms and no $H_2$ molecule can be formed. In almost simultaneous works, Heinola et al.[14] and Johnson and Carter,[15] have taken into account the temperature factor, and demonstrated that a vacancy only adsorbs five H atoms at room temperature and no molecules form as expected. For another point defect, substitutional He, Jiang et al.'s work[16] have demonstrated that it can also attract as many as 12 H atoms and no molecules form.

For extended defects like grain boundary and dislocation loop, the shortest dimension of their free volumes is smaller than a vacancy, but the other two dimensions are much larger. Seemingly, we may anticipate these defects being able to accommodate fewer H atoms in the short dimension than in a vacancy, and no $H_2$ molecule can sit along this direction. However, in the other two directions, i.e., in the grain boundary or dislocation loop plane, it is still possible to position the axes of $H_2$ molecules. A recent first-principles study by Zhou et al.[17] indicated that the $\Sigma 5(310)$ tilt GB in W can trap no more than two H atoms, and no $H_2$ molecules can form. However, H trapping by a vacancy, which is very easy to form at the GB, remains unknown.

To have a detailed knowledge of the local distribution of H near extended defects including grain boundaries and dislocation loops, we have exhaustively examined the segregation (trapping) of H to three kinds of traps, namely, $\Sigma 3(111)$ tilt grain boundary with and without an attached vacancy and an infinitely large dislocation loop in the (001) plane in W. The $\Sigma 3(111)$ GB has been studied by both experimentalists[18] and theorists.[19] The reason we chose a dislocation loop in (100) plane is in view of the reported experimental work in Ref. [6], in which the (100) surface of single crystalline W was electrically polished and irradiated by deuterium for a study of blistering. In addition, to assess the effect of H trapping on the cleavage along the GB or dislocation loop, we have also calculated the adsorption of H on the (100) and (111) planes of W.

Our DFT calculations were carried out using the Vienna *Ab initio* Simulation Package (VASP).[20] The electron-ion interaction was described using projector augmented wave method,[21,22] the exchange correlation potential using the generalized gradient approximation (GGA) in the Perdew-Burke-Ernzerhof form.[23] The lattice constant of bcc W was calculated to be 3.17 Å. For GB systems, we chose a Σ3(111) tilt GB of W using a 29-layer supercell imposing periodic boundary conditions [Fig. 1 (a)]. The slabs were separated by a 10 Å-thick vacuum avoid GB-GB interactions. GB(0) denotes the core of GB, and the other atomic sites are orderly labeled by numbers counted from the GB plane. One unit cell contains two mirror-symmetric 15-layer W atoms arranged at the <111> direction with one layer in common, which form a tilt GB in between. The thickness of such a cell is large enough to simulate a bulk-like environment in the center of the grain. For dislocation loops, one layer in a 16-layer supercell was removed to simulate a dislocation loop [Fig. 1 (b)]. Each layer has only one atom. The first four layers (two on each side) next to the loop were relaxed and W atoms in other layers were fixed. A (8×8×2) *k*-mesh within Monkhorst-Pack scheme and an energy cutoff of 250 eV were used in both systems. The geometry optimization for each system was continued until the forces on all the atoms were no larger than $10^{-3}$ eV/Å.

**FIG.1 (a) Side and top views of the computational cell used to model the Σ3 (111) tilt GB in bcc W. GB(0) denotes the interstitial site at the GB core. Atoms near the GB are numbered by the atomic layer counted from the GB plane. (b) Model of a dislocation loop in the (100) plane. Light green is the missing W layer.**

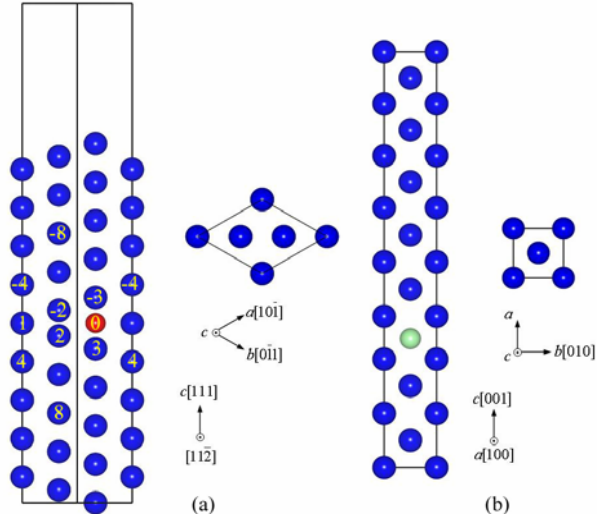

H atoms were added one by one to the pure Σ3(111) tilt GB. The first H prefers to stay close to GB(0) site at the center of the GB core (see Figure 2). The segregation (trapping energy), in reference to the H in a perfect bulk interstitial position, is -1.18 eV. And with zero-point energy (ZPE) corection, it is -1.31 eV. For the second and third H, they are -1.00 and -0.82 eV (-1.12 and -0.91 eV with ZPE). No more H can be trapped at the GB core even at zero temperature, and the trapped three constitute an equilateral triangle with a side length of 2.03 Å, much larger than H-H bondlength in $H_2$ (0.75 Å). Thus, we can conclude that no molecule can be formed at an ideal Σ3(111) GB in W, very similar to the Σ5(310) case.[17]

Grain boundaries are known to often serve as vacancy sinks. We have evaluated the formation energy of a vacancy at sites from GB(1) to GB(8). It is most stable and exothermic (-0.04 eV) at GB(2), an indication that the absence of W at GB(2) or GB(-2) is energetically favorable. This is a consequence of the strong repelling between GB(2) and GB(-2) at an ideal boundary structure. It is 3.66, 2.36, and 3.75 eV at the GB(1), GB(3), GB(4) sites, respectively. At GB(8) it is 3.20 eV, close to the value (3.30 eV) in a perfect bulk environment. Before implanting H atoms, we optimized the *c* axis, and this value was fixed upon introducing H atoms. Very interestingly, we find from the optimized geometry that upon removing of GB(2), all of the three H diffuse into the vacancy from the GB core (see Fig.2). On the other hand, if we introduce the vacancy to the GB first and then we add H atoms one by one into the vacancy at GB(2), when the number reaches three, we would obtain exactly the same configuration as above. This double check not only demonstrates the reliability of the first set of calculations, but also tells us that the distribution of trapped H atoms is independent of the order of the segregation of H and vacancy. Notice that, now, the GB(3) W is shifted up and the GB core region around GB(0) disappears. In fact, the GB structure becomes symmetric again with the GB(3) as the common layer. The positioning of the fourth H is rather critical. It can either push GB(3) up or down by equal chance. We have chosen the former configuration to continue the addition of forthcoming H atoms (Fig. 2). For each and every new coming H, we have examined many possible stable configurations, using knowledge gained from previous works.[13,16] The maximum number of H atoms trapped by a $\Sigma 3(111)$ GB combined with a vacancy is nine at zero temperature. We stress that no $H_2$ molecules are formed. The calculated trapping energy for each H is shown in Fig. 3. At room temperature, at least three H can be trapped in this GB.

FIG.2 The calculated geometry and isosurfaces of charge density near a Σ3(111) GB and (111) free surface of W with various number H atoms. Large and blue are W, small and red are H atoms. Yellow curves stand for the charge density isosurfaces of 0.02$e$ a.u.$^{-3}$.

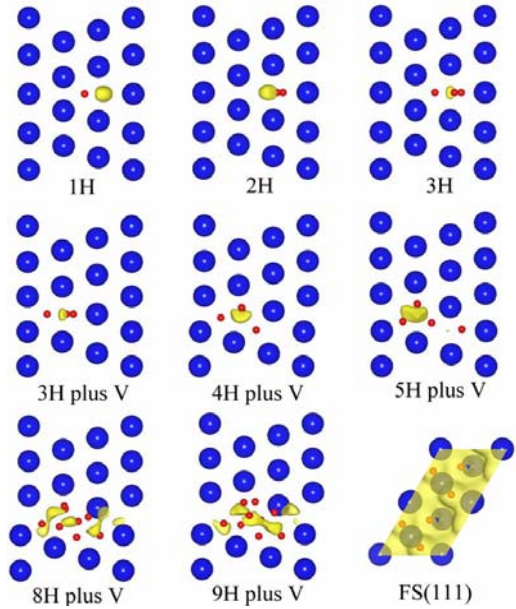

FIG.3 The calculated trapping energy when one H atom is attracted from inside a grain to the Σ3(111) GB combined with a vacancy in W, with and without zero-point energy correction. The trapping process is in a sequential manner.

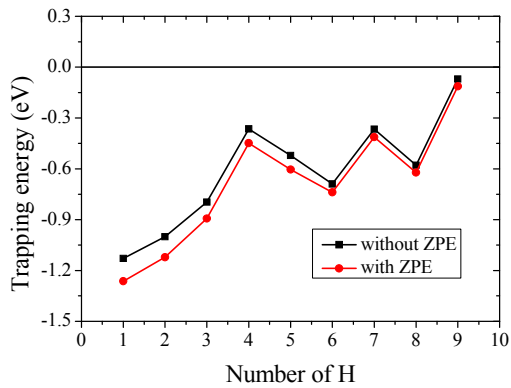

Rice Wang thermodynamic theory [24] is appropriate to describe the mechanism of impurity-induced embrittlement by the competition between dislocation crack blunting and brittle boundary separation. The potency of a segregation impurity in reducing the Griffith energy of a brittle boundary separation is a linear function of the difference in binding energies for that impurity at the GB and the free surface.[25, 26] To assess the effect of H accumulation at the GB on the decohesion across the boundary plane, we need to investigate the adsorption of H on W (111) surface, which is the eventual state of blistering or cleavage. Our DFT calculations show that totally three H atoms can be adsorbed within the (1×1) unit cell, and the binding energies are -1.63, -1.43, and -1.29 eV respectively. Corrected with ZPE, they become -1.72, -1.51, and -1.36 eV. In the absence of H, the cleavage energy of W along a $\Sigma 3(111)$ GB [with an associated vacancy at GB(2)] is calculated to be 5.41. With nine H segregated, the final state will be three H on each (111) surface and one and a half $H_2$ molecule. Now, the cleavage energy is as small as -0.54 eV. This means that with a high density of trapped H, the cleavage along $\Sigma 3(111)$ GB in W could become exothermic. It is noteworthy that there are chances, especially during damaging irradiation, for more than one vacancy to form at the GB (per unit cell), and the cleavage could become even more exothermic. A spontaneous formation of H blistering will then follow. At high temperature, on the other hand, fewer H can be trapped at the GB, the decohesion of the GB and the chance for H blistering will be reduced.

We note that with the model adopted here, it is not well defined for the internal stress induced by the H accumulation at the GB in a precise manner. Since we have fixed the slab thickness upon adding H to the GB, the atomic compressive force imposed on the top and bottom layers of the slab have been readily obtained. It is found that with only H trapped at the GB, the stress on the surface layer of the slab has already attained 10 GPa, which is believed to the magnitude needed for dislocation loop punching. Therefore, we might expect that accumulation of H at the GB is likely, to say the least, to punch dislocation loops in W.

**FIG.4 The calculated geometry and isosurfaces of charge density at a dislocation loop in the (100) plane and the (100) free surface. Large and blue are W, small and red are H. Yellow curves stand for the charge density isosurfaces of 0.02$e$ a.u.$^{-3}$.**

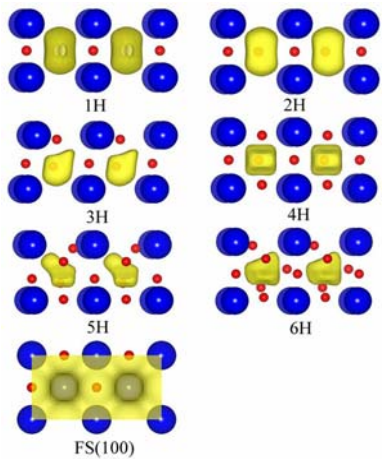

Next, we have performed a similar investigation on an infinitely large dislocation loop in the (001) plane in W. (see Fig.1b) Although the border of a dislocation loop cannot be counted by this model, the H-trapping capacity can be estimated reasonably to a large extent. As we did above, we optimized the $c$ axis of the spercell before adding H atoms one by one into the defect. The sequence of H introduction is depicted in Fig. 4. The first H atom was located at the distorted octahedral site, exactly in the plane of the missing atomic layer. The second one then occupied another octahedral site in the neighboring facet. The trapping energies of these two H atoms here is approximately equivalent to those on a (100) free surface. The third and fourth H went to a distorted octahedral site of the top and bottom facet. Upon introducing a fifth one, all these H atoms made a transform to occupy neighboring tetrahedral sites. At last, six H atoms occupied six facets, respectively. Our exhaustive search shows that totally six H atoms can be trapped in each square $a^2$ (a is the lattice constant) of this dislocation loop. Again, our DFT calculations demonstrated unambiguously that no $H_2$ molecules are formed. The calculated trapping energy for each H is shown in Fig. 5. The feature shown here is very similar to the case of H trapping by substitutional He[16] or a mono-vacancy,[12,13,14,15] as the underlying physics is vey similar.

**FIG.5 The calculated trapping energy when one H atom is attracted from a perfect W bulk environment to the inside of a dislocation loop, with and without zero-point energy correction. The trapping process is in a sequential manner.**

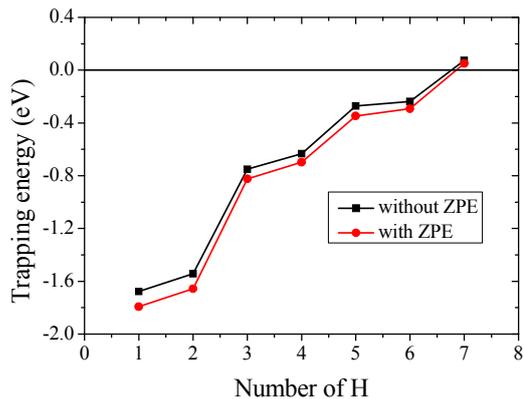

To assess the effect of H accumulation at the dislocation loop on the decohesion across the loop, we have investigated the adsorption of H on W (100) free surface, which is the eventual state after fracture. We find that only two H atoms can be absorbed within the (1×1) unit cell on the (100) surface. The distance between the two is 2.2Å and the binding energies are 1.74 and 1.76 eV, in reference to a H in the perfect bulk. Corrected with ZPE, the trapping energies both become -1.83 eV. With six H segregated, the final state of cleavage will be two H on each (100) surface and one $H_2$ molecules per (1×1) unit cell. Now, the cleavage energy is reduced from 1.80 eV in the absence of H to -1.96 eV with the help of H atoms. Actually, with four H segregated at high temperature, it could still be reduced to -0.60 eV. This means that with a high density of trapped H, the fracture along (100) dislocation loop could become strongly exothermic.

In conclusion, our density functional theory calculations on H trapping at the $\Sigma 3(111)$ tilt grain boundary and dislocation loop in (100) plane indicate that both of the extended defects can trap high density of H atoms. The trapped H weakens severely the cohesion across the boundary or loop plane and can therefore serve as precursors of H blistering. We observe no $H_2$ molecules are formed, supporting the speculation from experimentalist that the appearance of molecules is related to the formation of voids.[27] Furthermore, a comparison of decohesion effect introduced by H on the grain boundary and dislocation loop in W points to the relevance of free volume associated with defects. The larger the free volume, the stronger the embrittling effect of trapped H, and hence the more compelling force to drive blistering. As a consequence, one strategy to combat H blistering in metals could be reducing the free volume at grain boundaries by introducing segregants with small atomic size. For W, candidates might include Be, B, C, N, and O.

**ACKNOWLEDGMENTS**

The work was supported by the NSFC (Grant No. 50971029), NSFC-ANR (Grant No. 51061130558) and MOST (Grant No. 2009GB109004) of China. The calculations were performed on the Quantum Materials Simulator of USTB.